\documentclass[3p,times]{elsarticle}
\usepackage{textcomp}
\usepackage{amssymb}
\usepackage{amsthm}
\usepackage{amsmath}

\usepackage[figuresright]{rotating}

\usepackage{subfig}

\usepackage{graphicx}
\usepackage{dcolumn}
\usepackage{bm}
\usepackage[colorlinks = true,
linkcolor = blue,
urlcolor  = blue,
citecolor = blue,
anchorcolor = blue]{hyperref}
\usepackage{url}

\begin{document}
	
	\begin{frontmatter}
		
		\title{From Snaking to Isolas: A One-Active-Site Approximation in Discrete Optical Cavities}
		
		
		\author[au1,au1b]{R.\ Kusdiantara\corref{cor1}}
		\ead{rudy\_kusdiantara@itb.ac.id}	
		\author[au5,au3]{H.\ Susanto}
		\author[au2]{A. R. Champneys}	
		
		\address[au1]{Industrial and Financial Mathematics Research Group, Institut Teknologi Bandung, Jl.\ Ganesha No.\ 10, Bandung, 40132, Indonesia}
		\address[au1b]{Centre of Mathematical Modelling and Simulation, Institut Teknologi Bandung, Jl.\ Ganesha No.\ 10, Bandung, 40132, Indonesia}
		\cortext[cor1]{Corresponding author}
		\address[au5]{Department of Mathematics, Khalifa University, PO Box 127788, Abu Dhabi, United Arab Emirates}
		\address[au3]{Department of Mathematics, Faculty of Mathematics and Natural Sciences, Universitas Indonesia,\\ Gedung D Lt.\ 2 FMIPA Kampus UI Depok, 16424, Indonesia}
		\address[au2]{Department of Engineering Mathematics, University of Bristol, Bristol BS8 1UB, United Kingdom}

		\begin{abstract}
			
			We investigate time-independent solutions of a discrete optical cavity model featuring saturable Kerr nonlinearity, a discrete version of the Lugiato-Lefever equation. This model supports continuous wave (uniform) and localized (discrete soliton) solutions. Stationary bright solitons arise through the interaction of dark and bright uniform states, forming a homoclinic snaking bifurcation diagram within the Pomeau pinning region. As the system approaches the anti-continuum limit (weak coupling), this snaking bifurcation widens and transitions into $\subset$-shaped isolas. We propose a one-active-site approximation that effectively captures the system's behavior in this regime. The approximation also provides insight into the stability properties of soliton states. Numerical continuation and spectral analysis confirm the accuracy of this semianalytical method, showing excellent agreement with the full model.
			
		\end{abstract}
		
		
	\end{frontmatter}
	
	
	\section{Introduction}
	In recent years, there has been growing interest in localized patterns within nonlinear systems, particularly those exhibiting complex multiplicity, now commonly referred to as homoclinic snaking \cite{kozyreff2006asymptotics,burke2006localized,burke2007homoclinic,burke2007snakes,burke2012localized,lloyd2008localized,lloyd2017continuation,dawes2008localized,dawes2010emergence,beck2009snakes}. 
	For a comprehensive review of the wide-ranging applications of homoclinic snaking in continuous media, see \cite{knobloch2015spatial}.
	Although homoclinic snaking has predominantly been studied in spatially continuous systems, it also manifests in spatially discrete models \cite{peschel2004discrete,kusdiantara2017homoclinic,taylor2010snaking,chong2010variational,yulin2010discrete,yulin2010snake,susanto2018snakes}.
	
	In general, localized patterns, as opposed to isolated peaks or spots, form in systems exhibiting bistability between a background state and an excited or patterned state.
	In spatially continuous models, the patterned state is often represented as a periodic array known as "rolls," inspired by fluid convection problems. 
	In spatially discrete models, however, the excited state inherently possesses a length scale defined by the site separation. 
	Localized states are formed by pairs of forward and backward fronts that connect the excited and background states. Consequently, for fixed parameter values, stable localized patterns of arbitrarily large extent can emerge \cite{sakaguchi1996stable}.
	
	Following the pioneering work of Yves Pomeau \cite{pomeau1986front}, the parameter region where these fronts become pinned together is referred to as the ``pinning region''. 
	Homoclinic snaking \cite{woods1999heteroclinic} describes the bifurcation diagram that characterizes the morphogenesis of the simplest localized structures as a bifurcation parameter varies across the pinning region (see, for example,\ Fig.\ \ref{fig:snaking_two_cases} below).
	As the bifurcation parameter is varied, localized structures with increasing overall mass (or periodicity) are created through a sequence of folds. Each pair of folds introduces an additional roll or site within the excited region.
	In many contexts, each successive fold either stabilizes or destabilizes the solution branch. As a result, infinitely many stable pulses of arbitrary extent can be observed.
	Furthermore, the variational approximation, a well-known tool for approximating localized states governing homoclinic snaking in nonlinear systems, has also been studied in \cite{carretero2006multistable,chong2009multistable,chong2012validity}.
	
	In nonlinear optics, localized structures arising from homoclinic snaking have been experimentally observed in various systems, including driven optical systems \cite{purwins2001self}, 2D vertical-cavity surface-emitting lasers (VCSELs) \cite{tlidi2012delay}, vertical-cavity semiconductor optical amplifiers \cite{barbay2008homoclinic}, spatially forced systems \cite{haudin2011homoclinic}, and liquid crystal light-valves \cite{bortolozzo2009solitary}. Additional phenomena such as the existence of chimera-like states in one- and two-dimensional discrete Lugiato-Lefever models \cite{clerc2017chimera,clerc2020two}, and the coexistence of cavity solitons with different polarization states \cite{averlant2017coexistence} have also been reported.
	
	The precise structure of homoclinic snaking has been described in certain singular versions of the continuum limit using beyond-all-orders asymptotics \cite{kozyreff2006asymptotics,chapman2009exponential,dean2011exponential}, where the snaking region becomes exponentially narrow in the singular parameter.
	In conservative or variational problems, the center of the snaking region corresponds to the Maxwell point, a parameter at which the background and excited states possess equal energy \cite{pomeau1986front,hunt2000cellular,budd2001asymptotics}. 
	%
	This paper aims to investigate the opposite limit in spatially discrete systems: the weakly coupled, or anti-continuum, limit, where the pinning region reaches its maximum width.
	Our approach builds on earlier asymptotic techniques, utilizing variational or single-mode approximations, as presented in \cite{susanto2011variational,kusdiantara2017homoclinic}.
	While other rigorous approaches, such as the work of Beck {\em et al} \cite{beck2009snakes}, can prove the existence of localized structures and describe their bifurcation diagrams, our goal is to develop simplified methods capable of producing quantitative predictions.
	
	While our method is quite general, in this work, we focus on a discrete Ginzburg-Landau-type equation, specifically the discrete Lugiato-Lefever equation with saturable nonlinearity introduced by Yulin {\em et al.} \cite{yulin2008discrete,yulin2010discrete,yulin2010snake}, as a canonical model for optical cavities.
	This equation models light propagation in an array of weakly coupled optical waveguides \cite{egorov2003discrete,egorov2007subdiffractive,egorov2007does}. Saturation gives rise to families of both bright and gray solitons, exhibiting multistability as they develop internal shelves in the pinning region under both zero and finite losses \cite{yulin2008discrete}. 
	
	The primary aim of this report is to demonstrate how our new anti-continuum approximation performs on the example system introduced in \cite{yulin2010snake}. 
	Additionally, we provide further details on how the homoclinic snake breaks up into a series of {$\subset$-shaped} isolas as the pinning region widens in the weakly coupled limit \cite{moore2012frontiers,avitabile2012numerical}.
	
	The rest of the paper is organized as follows. In Section \ref{sec:unisol}, we discuss the spatially discrete governing equation and study its uniform solutions and their linear stability.  
	We discuss localized solutions, homoclinic snaking, and the	formation of {$\subset$-shaped} isolas in Section
	\ref{sec:loc_snake}.
	Section \ref{sec:pinning_anal} introduces our one-active-site approximation method to approximate the width of the pinning region.	We also compare the result with numerical computations in the section, where good agreement is obtained provided
	that the arrays are weakly coupled.	Finally, Section \ref{sec:conclusion} draws conclusions. 
	
	\section{Mathematical Model}\label{sec:unisol}
	In this study, we consider the one-dimensional lattice equation for a complex field $A_n\in \mathbb{C}$, $n\in \mathbb{Z}$:
	\begin{equation}
		i\partial_t A_n+\delta A_n +\frac{\alpha|A_n|^2}{1+|A_n|^2}A_n+c{\Delta}A_n = P,
		\label{eq:or_cavity}
	\end{equation}
	where ${\Delta}A_n=A_{n+1}+A_{n-1}-2A_n$.
	$A_n$ represents the amplitude of the $n$th identical optical resonator in a one-dimensional array \cite{yulin2010discrete}, $c\geq0$ denotes the strength of the nearest-neighbour coupling between oscillators.
	$P$ is the amplitude of an applied optical pump field (real-valued and independent of $n$), which is our control/bifurcation parameter. 
	Re($\delta$)=$\delta_r$ represents the detuning of the pump frequency from the resonant frequency of the oscillators. 
	Re($\alpha$) indicates the strength of the Kerr effect of the intensity-dependent refractive index.
	Im($\delta$)=$\delta_i$  and Im($\alpha$) are linear and nonlinear loss terms, respectively.	Using the results of \cite{yulin2010discrete}, here we focus on two cases of parameter values, which yield rich dynamics and represent the general snaking behaviors in the discrete optical cavities, i.e.,
	\begin{itemize}
		\itemsep0em 
		\item Case 1: $
		\delta = -9.2+i \quad\textrm{and}\quad \alpha = 10,$
		\item Case 2: $
		\delta = 4+i \quad\textrm{and}\quad \alpha =-10.$
	\end{itemize}
	
	\begin{figure}[thbp!]
		\centering
		\subfloat[\,Case 1]{\includegraphics[scale=.5]{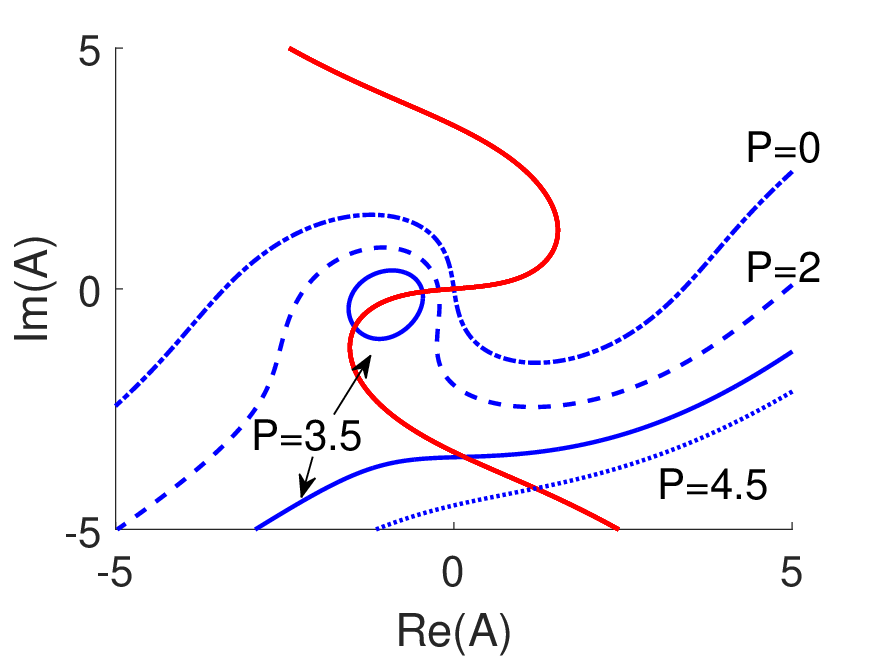}
			\label{subfig:alpha_10_delta_-9_2_1i_contour}}
		\subfloat[\,Case 2]{\includegraphics[scale=0.5]{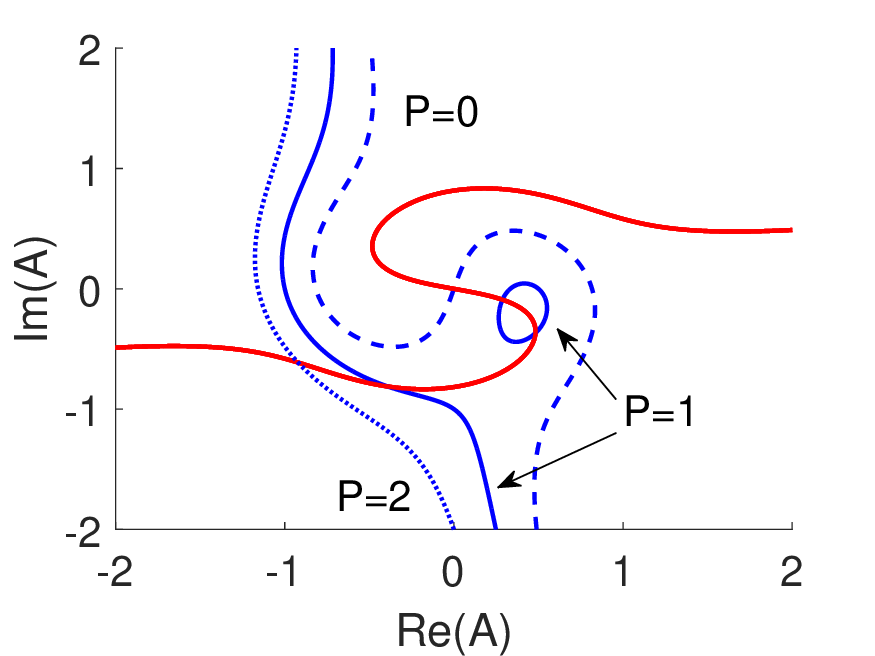}
			\label{subfig:alpha_-10_delta_4_1i_contour}}
		\caption{{(a), (b) Contour plots of the algebraic equation \eqref{eq:cavity_unisol} at several $P$. 
				The blue solid/dashed/dotted lines represent the real parts of the equations $Re(F_u)$, while the red solid lines are for the imaginary parts $Im(F_u)$. 
				Intersections between the blue and red curves represent uniform solutions of \eqref{eq:cavity_unisol}. 
				The red curve from Im$(F_u)=0$ is independent of the optical pump field $P$ because $P\in\mathbb{R}$.}
		}
		\label{fig:uniform_cav}
	\end{figure}
	\begin{figure}[thbp!]
		\centering
		\subfloat[\,Case 1]{\includegraphics[scale=0.46]{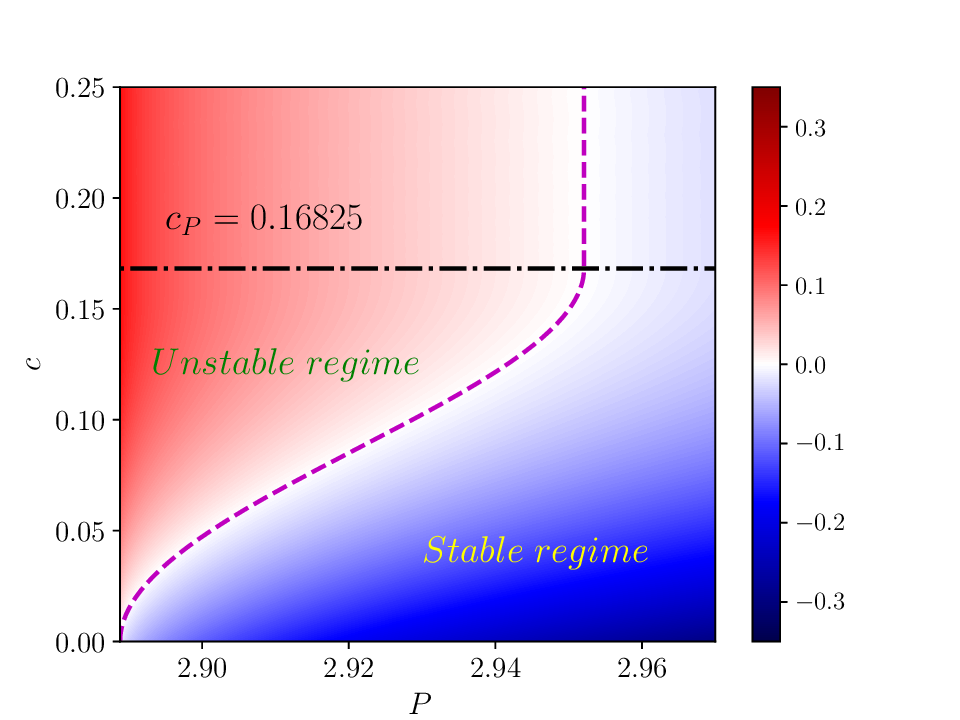}\label{subfig:stability_shift_case1_cav}}\quad
		\subfloat[\,Case 2]{\includegraphics[scale=0.46]{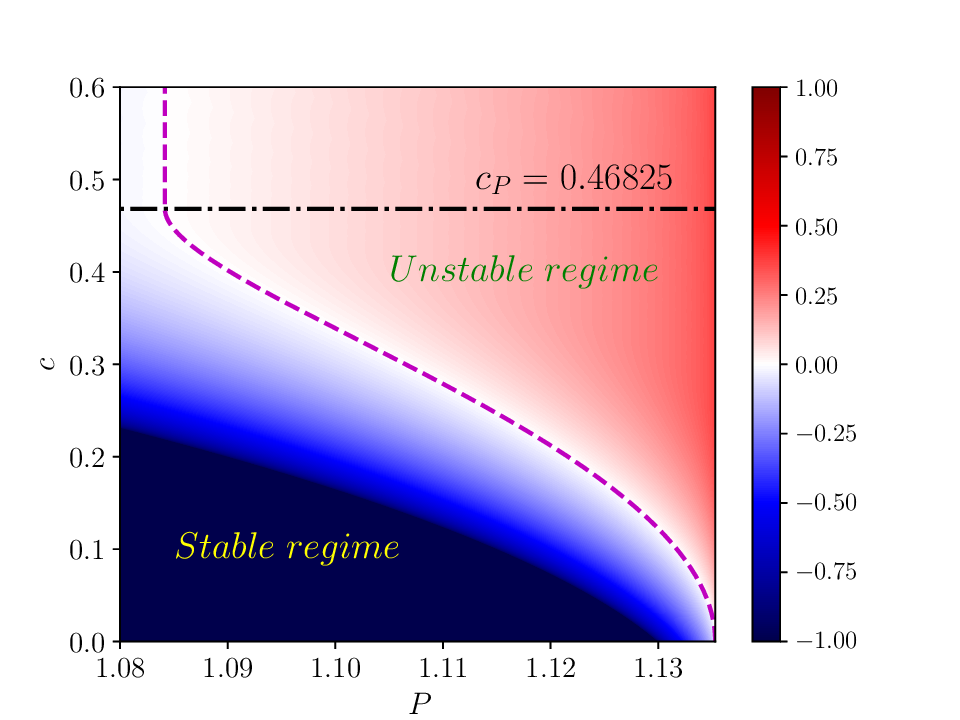}\label{subfig:stability_shift_case2_cav}}
		\caption{{Spectrum of largest eigenvalue around $P_1$ and $P_0$ for case 1 and 2, respectively}. 
			The (In)stability {point} of the uniform solutions for case 1 and 2 when $c$ is being varied {which is shown by dashed magenta lines}.}
		\label{fig:stability_shift_cav}		
	\end{figure}
	We consider the time-independent solution of equation \eqref{eq:or_cavity}, i.e.,
	\begin{equation}
		%
		{\delta A_n +\frac{\alpha|A_n|^2}{1+|A_n|^2}A_n+c{\Delta}A_n = P.}
		\label{eq:cavity_ti}
	\end{equation}
	Once a solution $A_n=\tilde{A}_n=\tilde{x}_n+i\,\tilde{y}_n$ is obtained, its	linear stability is then determined by writing
	\begin{equation}
		\begin{array}{rcl}
			x_n=\tilde{x}_n + {\epsilon}\hat{x}_ne^{\lambda t},\,
			y_n=\tilde{y}_n + {\epsilon}\hat{y}_ne^{\lambda t}.
		\end{array}
		\label{eq:anz_cavity_ri}
	\end{equation}
	Substituting \eqref{eq:anz_cavity_ri} into \eqref{eq:or_cavity}, linearising about  {$\epsilon=0$} and splitting the real and imaginary part of the resulting equations, we obtain the linear eigenvalue problem
	\begin{equation}
		\lambda\left(
		\begin{array}{c}
			\hat{x}_n\\
			\hat{y}_n
		\end{array}
		\right)=\mathcal{L}
		\left(
		\begin{array}{c}
			\hat{x}_n\\
			\hat{y}_n
		\end{array}
		\right),
	\end{equation}
	where
	\begin{equation}
		\mathcal{L}=\left(
		\begin{array}{cc}
			-\delta_i-m_{11}&-\delta_r -c\Delta-m_{12}\\
			\delta_r+c\Delta-m_{21}&-\delta_i-m_{22}
		\end{array}
		\right)
		\label{eq:evp_cav}
	\end{equation}	
	is the linear differential operator of Eq.\ \eqref{eq:or_cavity} and
	\begin{equation}
		\begin{array}{ccl}
			m_{11} &=&\displaystyle\frac{2\alpha \tilde{x}_n\tilde{y}_n}{\left(1+\tilde{x}_n^2+\tilde{y}_n^2\right)^2},\quad m_{12} =\displaystyle\frac{\alpha\left[\left(\tilde{x}_n^2+\tilde{y}_n^2\right)^2+\left(3\tilde{y}_n^2+\tilde{x}_n^2\right)\right]}{\left(1+\tilde{x}_n^2+\tilde{y}_n^2\right)^2},\\
			m_{21} &=&\displaystyle\frac{-\alpha\left[\left(\tilde{x}_n^2+\tilde{y}_n^2\right)^2+\left(3\tilde{x}_n^2+\tilde{y}_n^2\right)\right]}{\left(1+\tilde{x}_n^2+\tilde{y}_n^2\right)^2},\quad
			m_{22}=\displaystyle\frac{-2\alpha \tilde{x}_n\tilde{y}_n}{\left(1+\tilde{x}_n^2+\tilde{y}_n^2\right)^2}.
		\end{array}
		\label{eq:m_eq_cav}
	\end{equation}
	A solution is said to be stable when {Re$(\lambda)\leq0$} for all the eigenvalues and unstable otherwise.
	
	\begin{figure}[tbp!]
		\centering
		\subfloat[\,Case 1]{\includegraphics[scale=0.45]{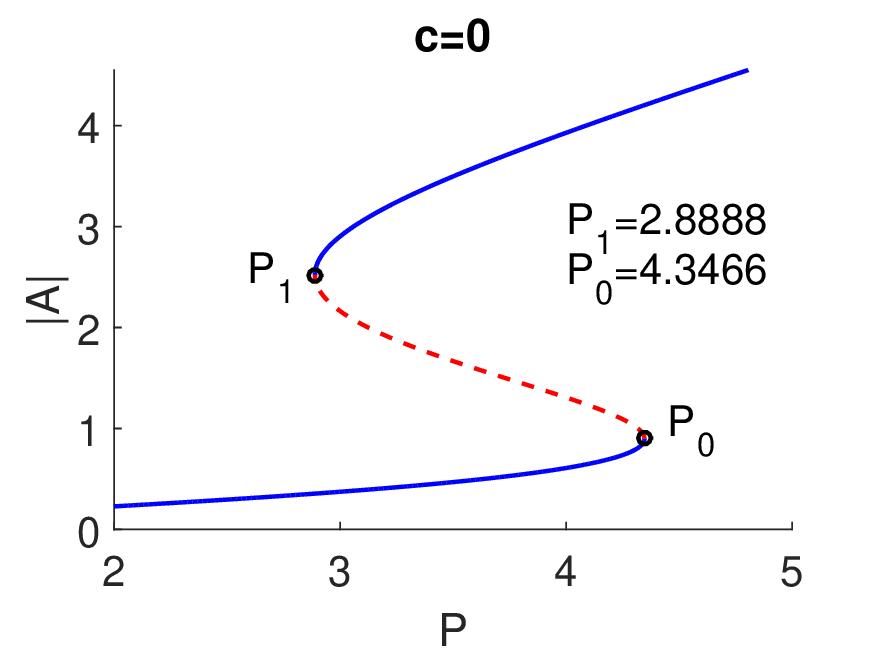}\label{subfig:case1_c_0}}
		\subfloat[\,Case 2]{\includegraphics[scale=0.45]{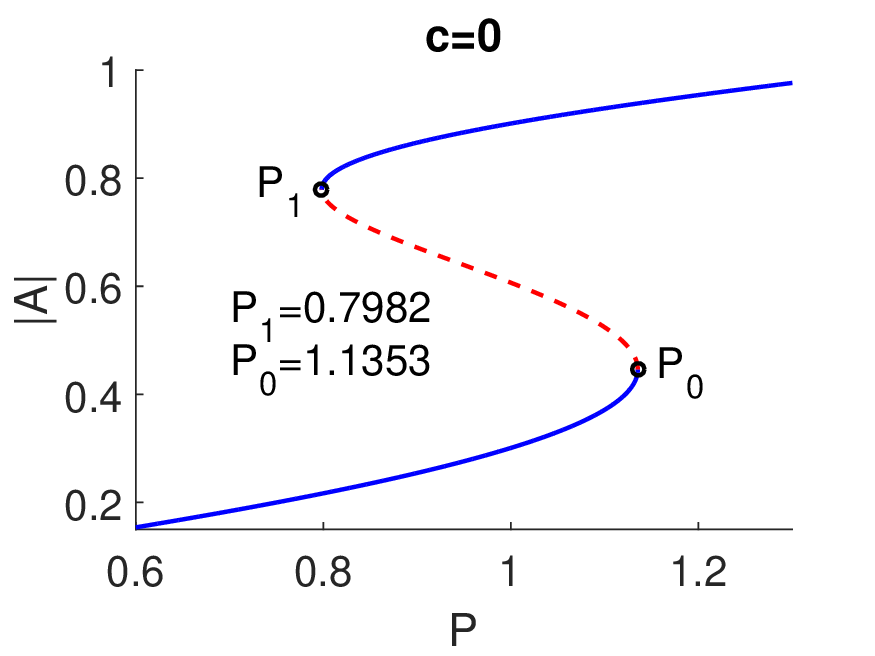}\label{subfig:case2_c_0}}\\
		\subfloat[\,Case 1]{\includegraphics[scale=0.45]{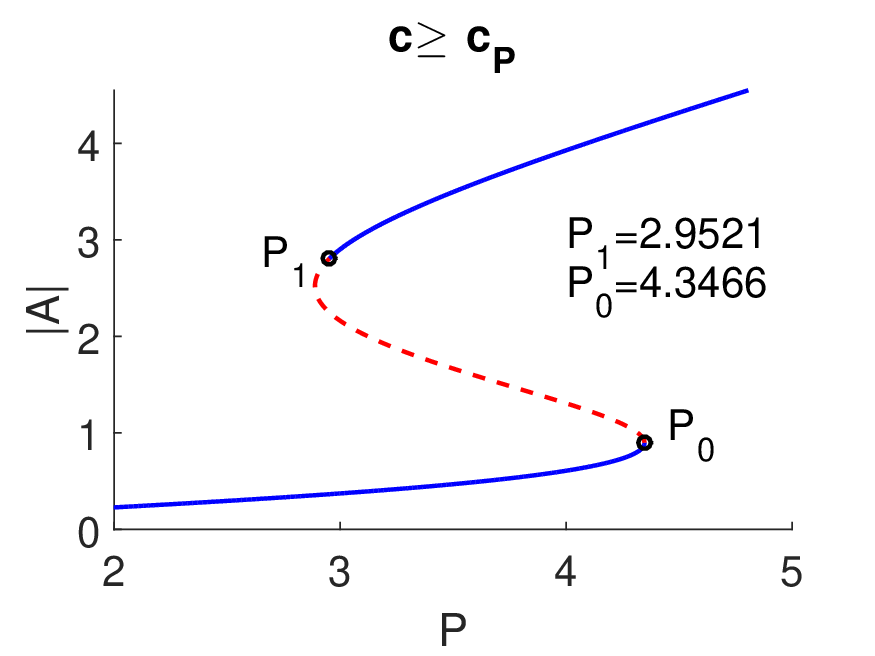}\label{subfig:case1_c_large_zoom}}	
		\subfloat[\,Case 2]{\includegraphics[scale=0.45]{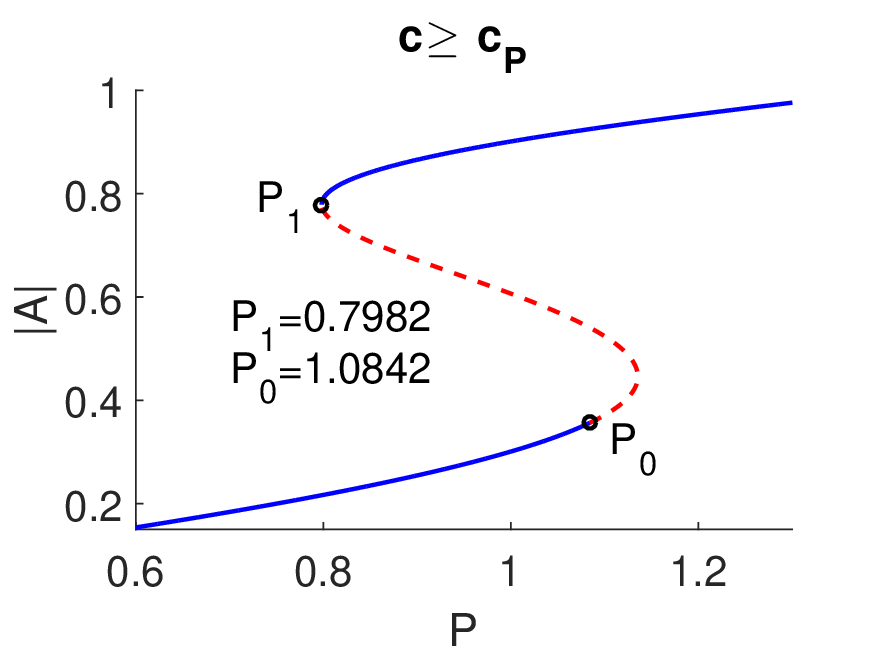}\label{subfig:case2_c_infty_zoom}}
		\caption{Bifurcation diagrams of the uniform solutions and their linear stability. 
			The blue solid and red dashed lines represent stable and unstable solutions.
			Note that there is a stability shift past the turning points as $c$ increases.
		}
		\label{fig:case_unisol}
	\end{figure}	
	\subsection{Uniform solutions}\label{subsec:unisol_stable}
	
	Equation \eqref{eq:or_cavity} has uniform solutions (homogeneous states) $A_n(t)=A$ satisfying
	\begin{equation}
		F_u\left(A\right)=\delta A +\frac{\alpha\,|A|^2}{1+|A|^2}A-P=0. \label{eq:cavity_unisol}
	\end{equation}
	
	Figure~\ref{fig:uniform_cav} illustrates the nullclines of the real and imaginary parts of $F_u$ as described in Eq.~\eqref{eq:cavity_unisol}, plotted for several values of the bifurcation parameter $P$. The intersections between the blue (real part) and red (imaginary part) curves correspond to uniform solutions of Eq.~\eqref{eq:cavity_unisol}. 
	Depending on the value of $P$, the system can have either one or three distinct uniform solutions. Specifically, for certain ranges of $P$, three solutions exist, and this interval is bounded by turning points where two of the solutions coalesce. These turning points mark the boundaries of the region in $P$ where multistability occurs.
	{We present in Fig.\ \ref{fig:case_unisol} the roots as a function of $P$. }
	
	To determine the linear stability of the uniform solutions, one has $\hat{x}_n=\hat{y}_n=e^{ikn}$, where $k$ is the wave number of the perturbation, from which we obtain the dispersion relation
	\begin{eqnarray}
		\lambda(k)=-\delta_i-\frac{1}{2}\left(m_{11}+m_{22}\right)\pm\frac{1}{2}\sqrt{\Gamma},
		\label{eq:dis_rel_cav}
	\end{eqnarray}
	where
	\begin{eqnarray*}
		\Gamma&=&-8\cos(k)c\left[2\cos(k)c+m_{12}-m_{21}-4c+2\delta_r\right)]-8c\left(2c-2\delta_r-m_{12}+m_{21}\right)+\left(m_{11}-m_{22}\right)^2\\
		&&-4\left(\delta_r-m_{21}\right)\left(m_{12}+\delta_r\right).
	\end{eqnarray*}
	A uniform solution is said to be stable when $\lambda(k)\leq 0$ for $\forall k\in \mathbb{R}$ and unstable when $\exists k $ such that $\lambda(k)>0$.
	The maximum of the spectrum \eqref{eq:dis_rel_cav} is attained at
	\begin{eqnarray}
		k=\pm\left\{
		\begin{array}{cc}
			\pi&,\ c< c_P\\
			\displaystyle\arccos\left(\frac{4c-m_{12}+m_{21}-2\delta_r}{4c}\right)&,\ c\geq c_P\\
		\end{array}
		\right.
		\label{eq:wave_num_cav}
	\end{eqnarray}
	where
	\begin{equation*}
		c_P = \frac{1}{8}\left(m_{12}-m_{21}-2\delta_r\right){,}
	\end{equation*}
	which is a parameter threshold for the {stability change point of} the uniform solution, see Fig.\ \ref{fig:stability_shift_cav}.
	
	
	Figures~\ref{subfig:case1_c_0} and \ref{subfig:case1_c_large_zoom} display the bifurcation diagrams of the uniform solutions for case 1. The point $P_1$, where the uniform solution transitions between stability and instability, shifts either to the right or left depending on the value of $c$. 	
	For $c < c_P$ (weakly coupled), the eigenvalue from Eq.~\eqref{eq:dis_rel_cav}, which determines the linear stability, varies with $c$. Specifically, $P_1$ shifts to the left as $c$ decreases, approaching the turning point. When $c = 0$ (uncoupled), $P_1$ is located at the leftmost turning point. 	
	In contrast, for $c \geq c_P$ (strongly coupled), the position of $P_1$ becomes independent of $c$.

	
	A similar mechanism occurs in case 2, where the stability change happens at $P_0$. This can be observed in Figs.~\ref{subfig:case2_c_0}, \ref{subfig:case2_c_infty_zoom}, and \ref{subfig:stability_shift_case2_cav}. 
	For case 2, as $c$ decreases, $P_0$ shifts to the right, indicating that the system becomes unstable at higher bifurcation parameter values. Conversely, as $c$ increases, $P_0$ shifts to the left, meaning the stability boundary moves towards lower values of $P$. 
	It is important to note that for both case 1 and case 2, there is a range of the bifurcation parameter where bistability occurs, specifically for $P_1 \leq P \leq P_0$. In this region, the system exhibits two stable uniform solutions, depending on the initial conditions or perturbations.

	\begin{figure*}[htbp!]
		\centering
		\subfloat[\,Case 1]{\includegraphics[scale=0.45]{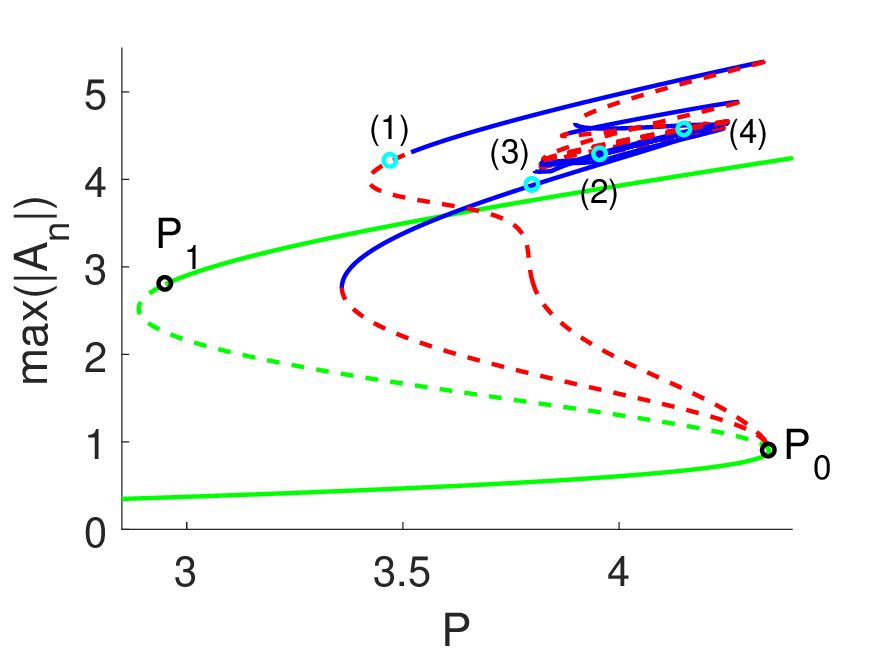}
			\label{subfig:alpha_10_delta_-9_2_1i_c_1}}
		\subfloat[\,Case 2]{\includegraphics[scale=0.45]{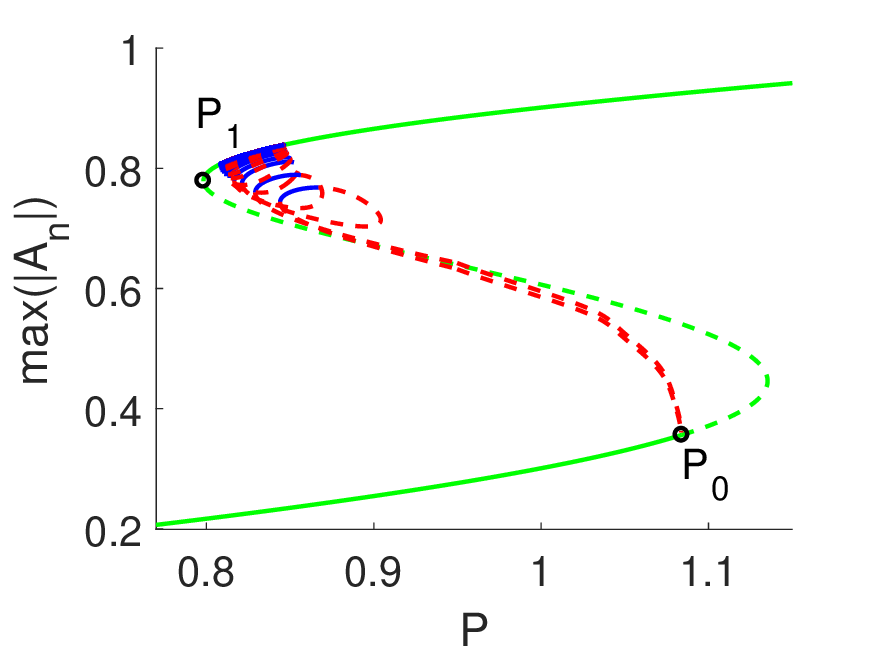}
			\label{subfig:alpha_-10_delta_4_1i_c_1}}\\
		\subfloat[\,Case 1]{\includegraphics[scale=0.45]{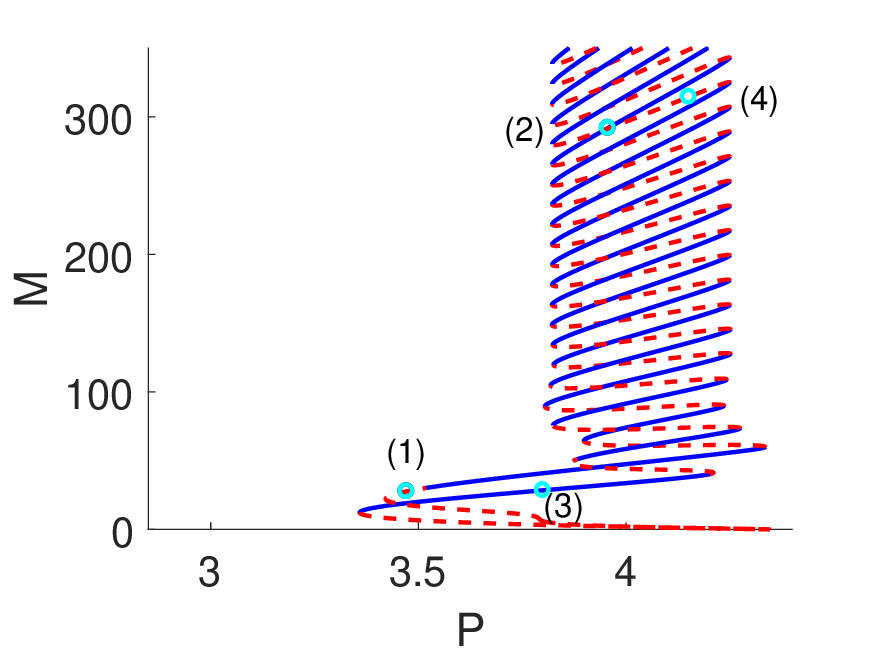}
			\label{subfig:alpha_10_delta_-9_2_1i_c_1_s}}
		\subfloat[\,Case 2]{\includegraphics[scale=0.45]{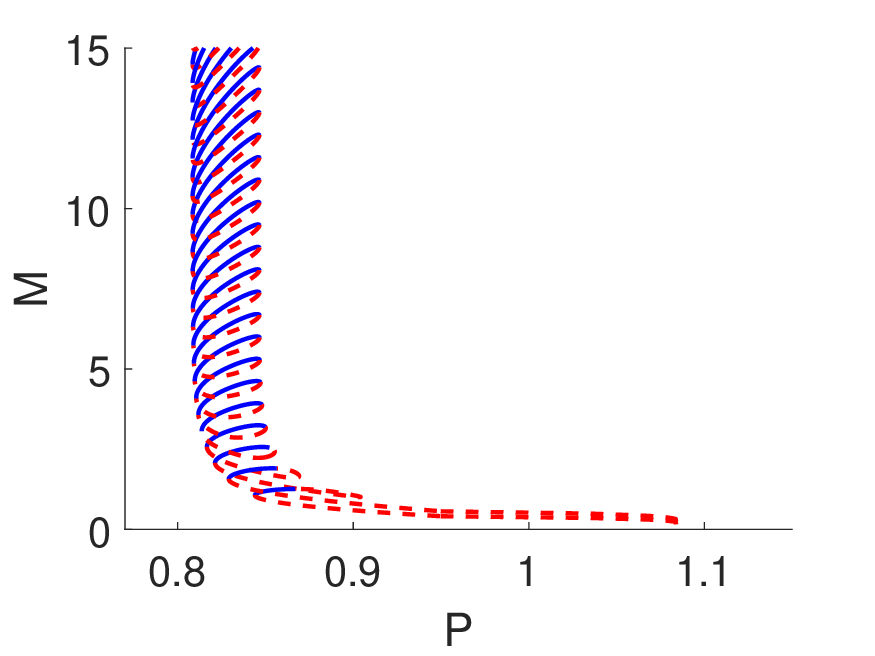}
			\label{subfig:alpha_-10_delta_4_1i_c_1_s}}
		\caption{Bifurcation diagrams for case 1 and 2 for $c=0.5$. The blue solid and red dashed lines represent stable and unstable {non-uniform} solutions, respectively{, i.e., homoclinic snaking}. The green solid/dashed line represents a uniform solution, {respectively stable and unstable}.
			The corresponding solutions at the indicated points in panels (a) and (c) are plotted in Fig.\ \ref{fig:onsite_offsite_several_P}.
		}
		\label{fig:snaking_two_cases}
	\end{figure*}
	\begin{figure*}[htbp]
		\centering
		\subfloat[]{\includegraphics[scale=0.45]{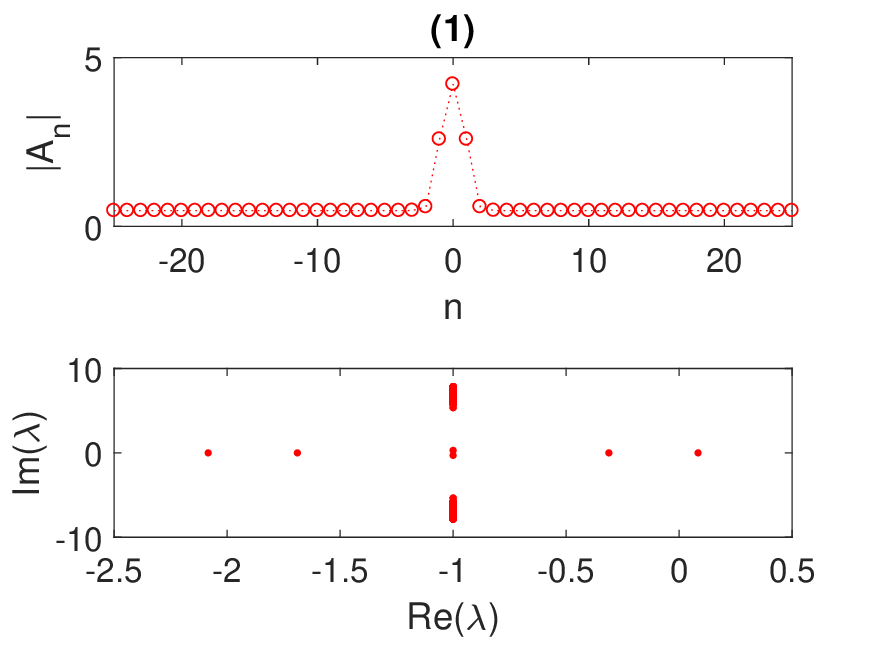}\label{subfig:prof_1_case1}}
		\subfloat[]{\includegraphics[scale=0.45]{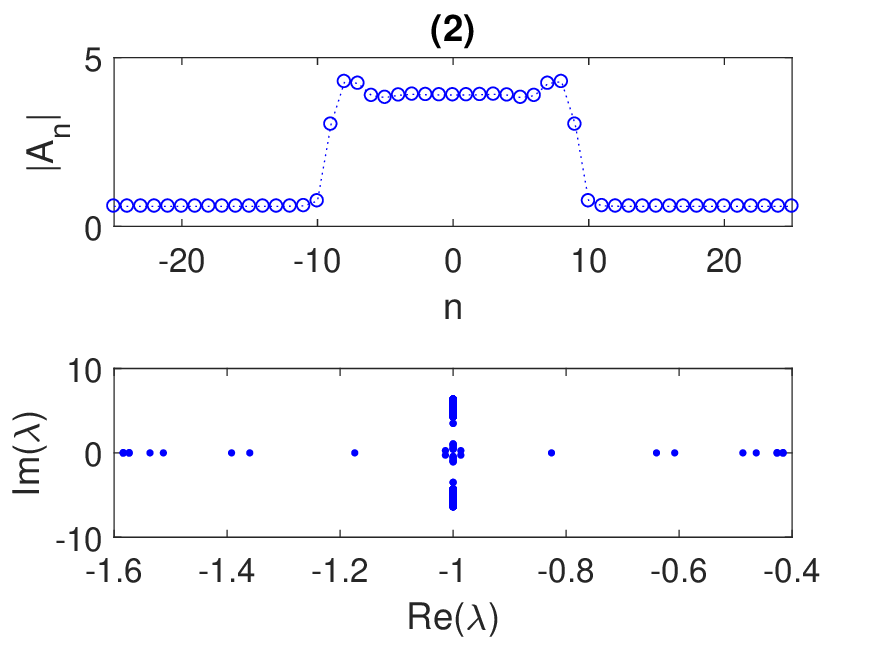}\label{subfig:prof_2_case1}}\\
		\subfloat[]{\includegraphics[scale=0.45]{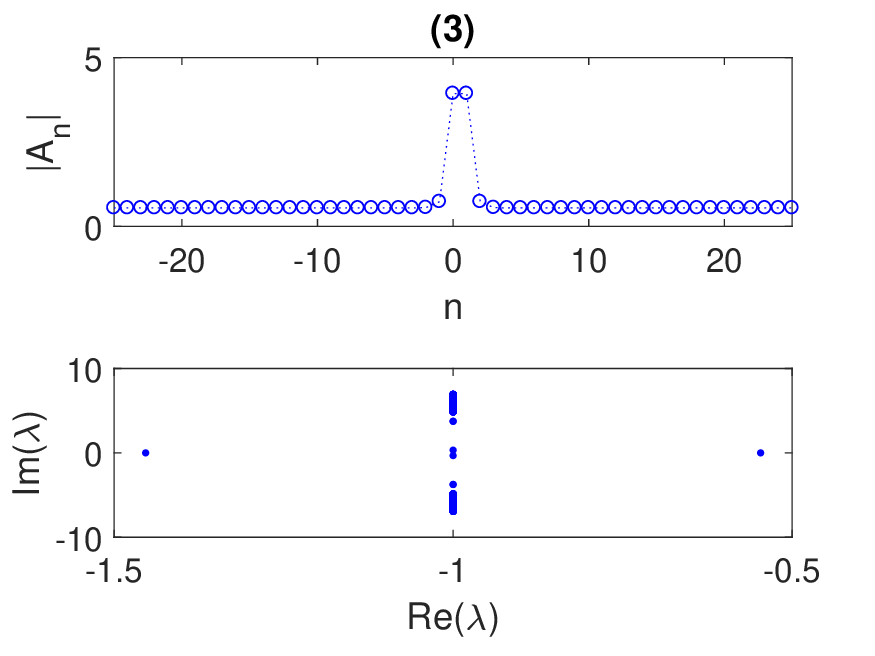}\label{subfig:prof_3_case1}}
		\subfloat[]{\includegraphics[scale=0.45]{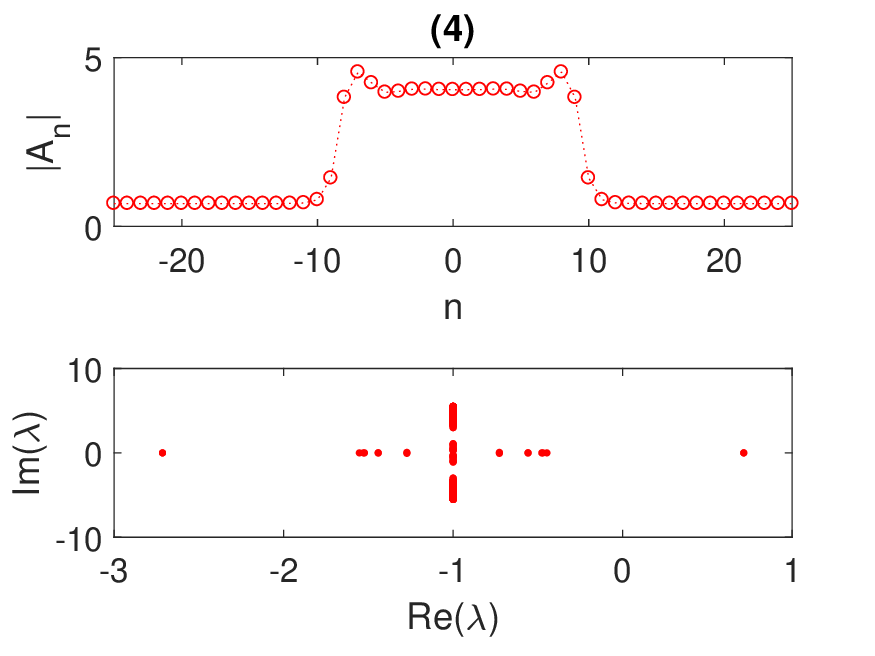}\label{subfig:prof_4_case1}}
		\caption{Plot of the on-site and off-site solution profiles on the bifurcation diagram in Figs.\ \ref{subfig:alpha_10_delta_-9_2_1i_c_1} and \ref{subfig:alpha_10_delta_-9_2_1i_c_1_s} and their spectrum in the complex plane. (a)-(b) and (c)-(d) depict on-site and off-site solutions, respectively.
			{The blue and red colors represent the stable and unstable solutions, respectively.}
		}
		\label{fig:onsite_offsite_several_P}
	\end{figure*}	
	\section{Localised solutions and Snaking}\label{sec:loc_snake}
	
	The discrete optical cavity equation \eqref{eq:or_cavity} admits localized solutions, specifically on-site and off-site solitons, which bifurcate from the uniform solutions at the bifurcation point $P_0$. 	
	These localized structures are formed by combining two stable uniform states, referred to as the "upper" and "lower" branches. The transition between these branches occurs through the interaction of fronts connecting the different uniform states, effectively creating a localized region where the two states coexist back-to-back. This type of solution represents a stable configuration where localized patterns emerge from the interaction of these uniform states.
	
	\subsection{Snaking}\label{subsec:snake}
	By applying numerical continuation, specifically the pseudo-arclength method~\cite{keller1987lectures}, for varying $P$, one can obtain bifurcation diagrams for both on-site and off-site solutions, as shown in Fig.~\ref{fig:snaking_two_cases}.
	To analyze these solutions, we introduce the ``norm'', defined as:
	\begin{eqnarray}
		M = \sum_n \left( |A_n - A_\infty|^2 \right),
	\end{eqnarray}
	which we refer to as the soliton mass~\cite{yulin2010discrete,yulin2010snake}. This norm quantifies the deviation of the localized soliton solution from the uniform background state $A_\infty$.
	By plotting the bifurcation diagrams in terms of the soliton mass, the homoclinic snaking behavior for both on-site and off-site solutions becomes evident, as shown in Figs.~\ref{subfig:alpha_10_delta_-9_2_1i_c_1} and \ref{subfig:alpha_-10_delta_4_1i_c_1}. The snaking structure, with alternating stable and unstable branches, is visible in the mass plot for both types of localized solutions, as highlighted in Figs.~\ref{subfig:alpha_10_delta_-9_2_1i_c_1_s} and \ref{subfig:alpha_-10_delta_4_1i_c_1_s}.
	
	\begin{figure}[t!]
		\centering
		\includegraphics[scale=0.75]{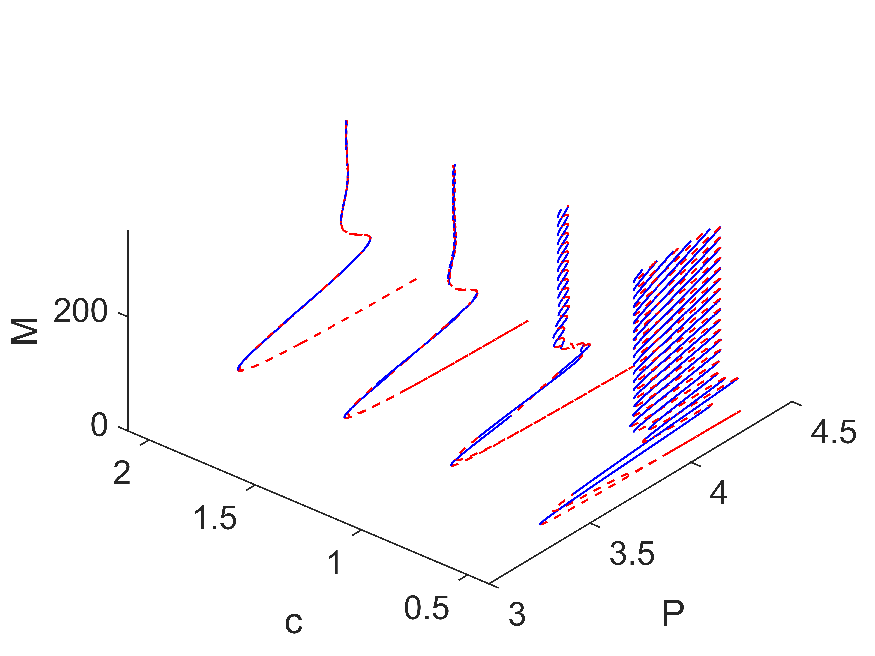}
		\caption{Bifurcation diagrams of case 1 for varying $c$. 
			{The narrowing of the snaking due to the coupling strength changes.}}
		\label{fig:several_snaking_cav}
	\end{figure}
	
	Figure~\ref{fig:onsite_offsite_several_P} presents the on-site and off-site localized solutions corresponding to the bifurcation diagram in Fig.~\ref{subfig:alpha_10_delta_-9_2_1i_c_1_s} for several values of the bifurcation parameter $P$. 
	As $P$ varies, the norm $M$, representing the soliton mass, increases. This corresponds to the "upper" state of the localized solution progressively invading the "lower" state, leading to a larger localized region. 
	Figure~\ref{fig:several_snaking_cav} shows the bifurcation diagrams illustrating the snaking behavior for different values of the coupling strength $c$ in case 1. The pinning region widens as the coupling strength $c$ decreases (i.e., the system becomes more weakly discrete). This indicates a broader range of parameter values where stable localized solutions exist. 
	In contrast, as $c \to \infty$ (strong coupling, approaching the continuum limit), no snaking behavior is observed~\cite{fiedler1996discretization,bramburger2020localized,bramburger2020spatially}. Furthermore, in the continuum limit, the on-site and off-site localized solutions merge into a single type of solution, erasing the distinction between them.
	
	\begin{figure*}[htbp]
		\centering
		\subfloat[]{\includegraphics[scale=0.45]{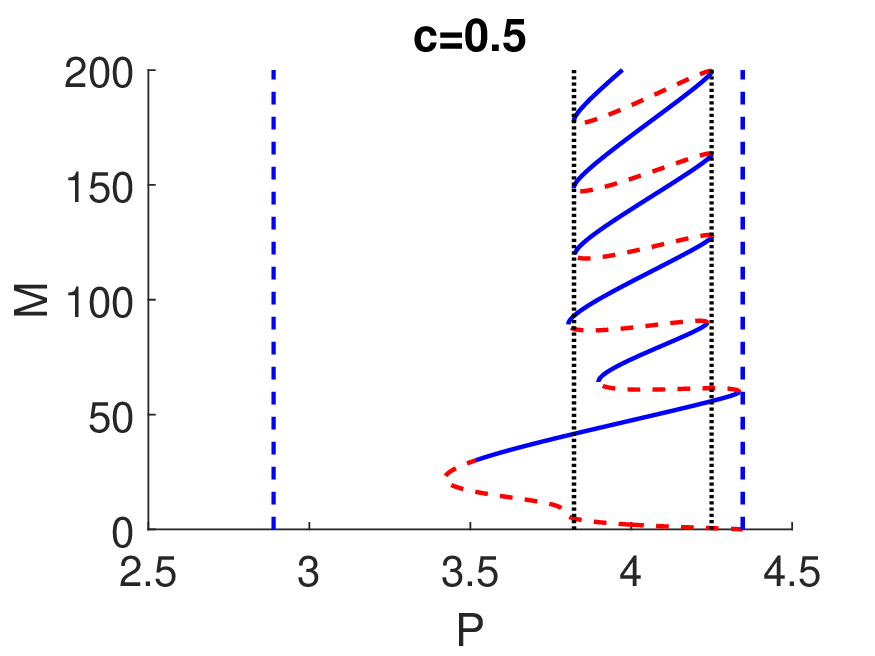}\label{subfig:alpha_10_delta_-9_2_1i_c_0_5}}
		\subfloat[]{\includegraphics[scale=0.45]{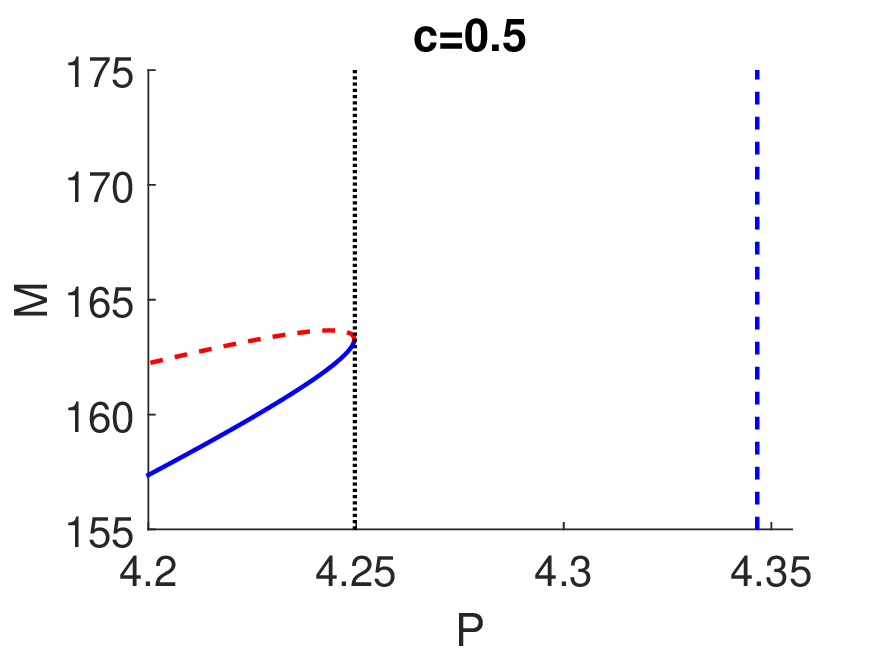}\label{subfig:alpha_10_delta_-9_2_1i_c_0_5_zoom}}\\
		\subfloat[]{\includegraphics[scale=0.45]{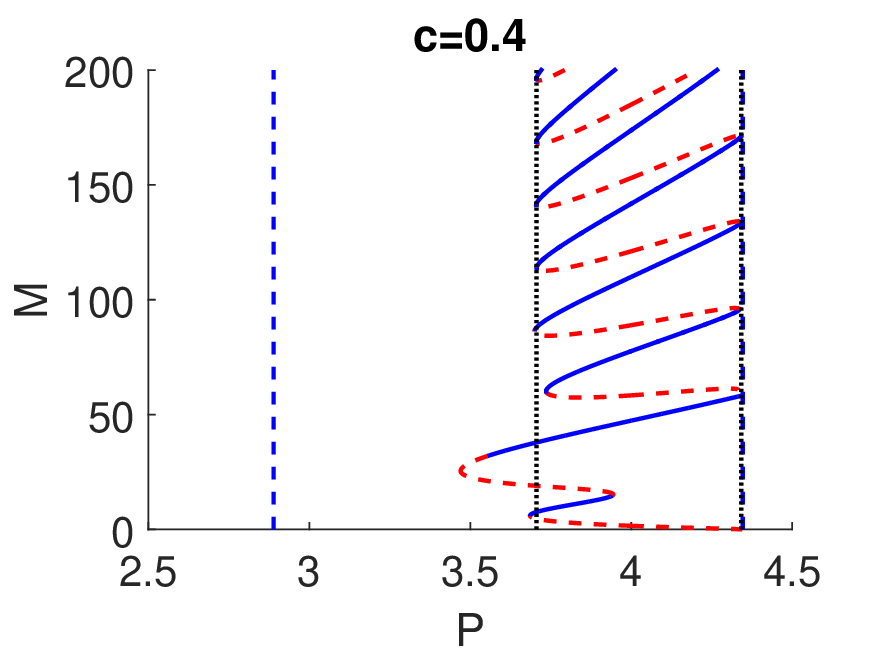}\label{subfig:alpha_10_delta_-9_2_1i_c_0_4}}
		\subfloat[]{\includegraphics[scale=0.45]{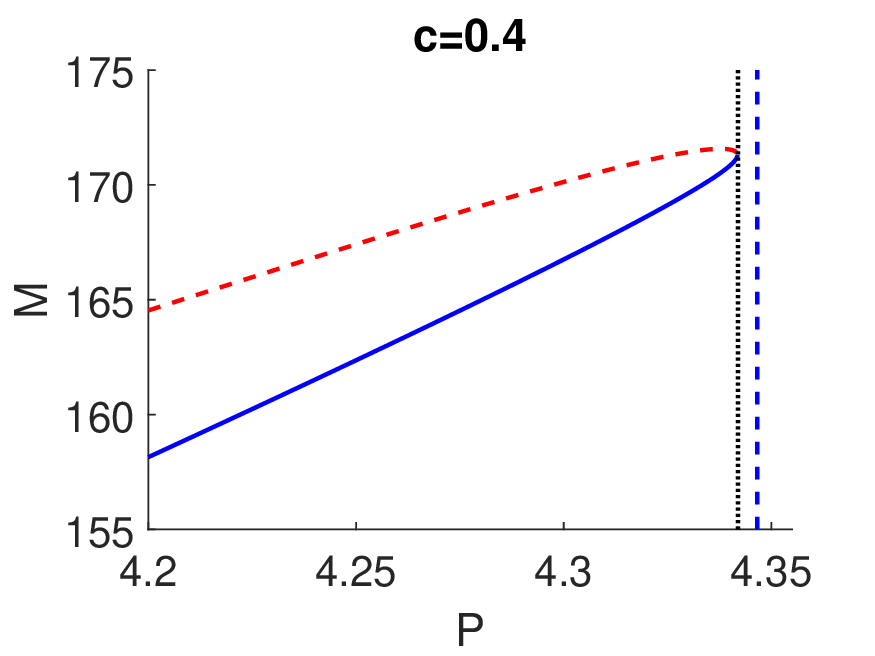}\label{subfig:alpha_10_delta_-9_2_1i_c_0_4_zoom}}\\
		\subfloat[]{\includegraphics[scale=0.45]{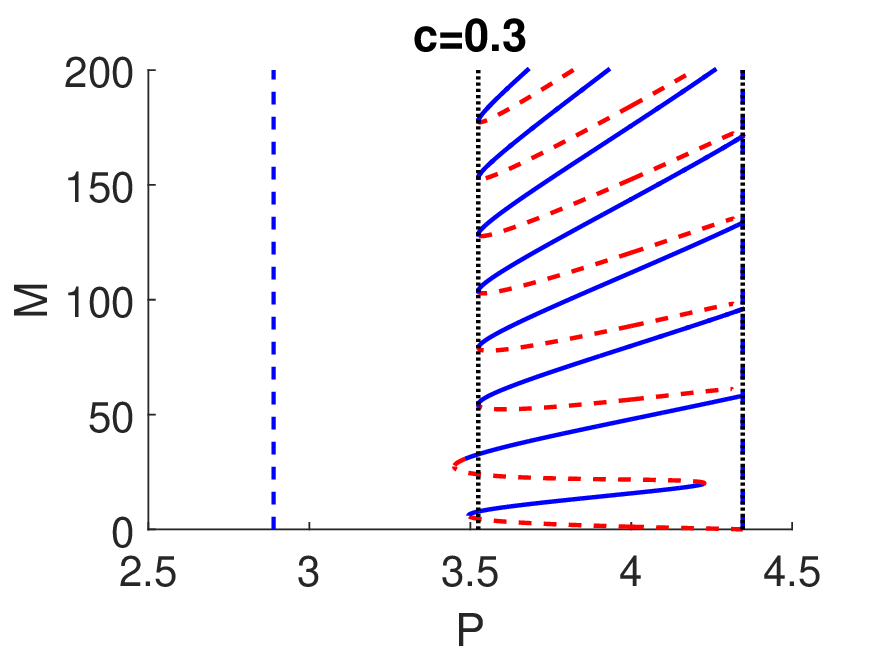}\label{subfig:alpha_10_delta_-9_2_1i_c_0_3}}
		\subfloat[]{\includegraphics[scale=0.45]{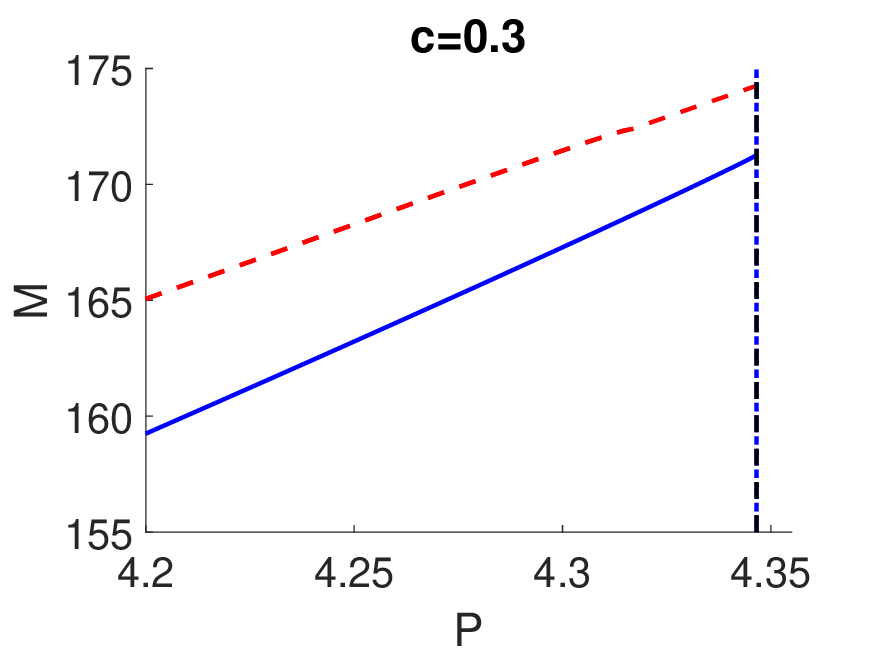}\label{subfig:alpha_10_delta_-9_2_1i_c_0_3_zoom}}
		\caption{The occurence of {$\subset$-shaped} isolas for case 1 (on-site solutions) when $c$ is being varied.}
		\label{fig:C_isolas_appear}
	\end{figure*}
	
	\begin{figure*}[htbp]
		\centering
		\subfloat[]{\includegraphics[scale=0.45]{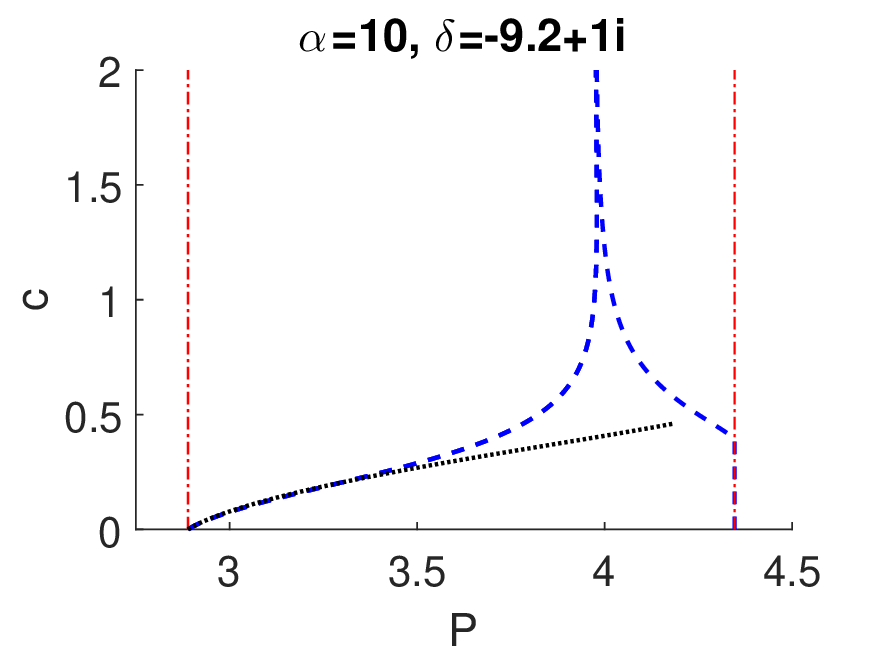}\label{subfig:1_P_vs_c}}
		\subfloat[]{\includegraphics[scale=0.45]{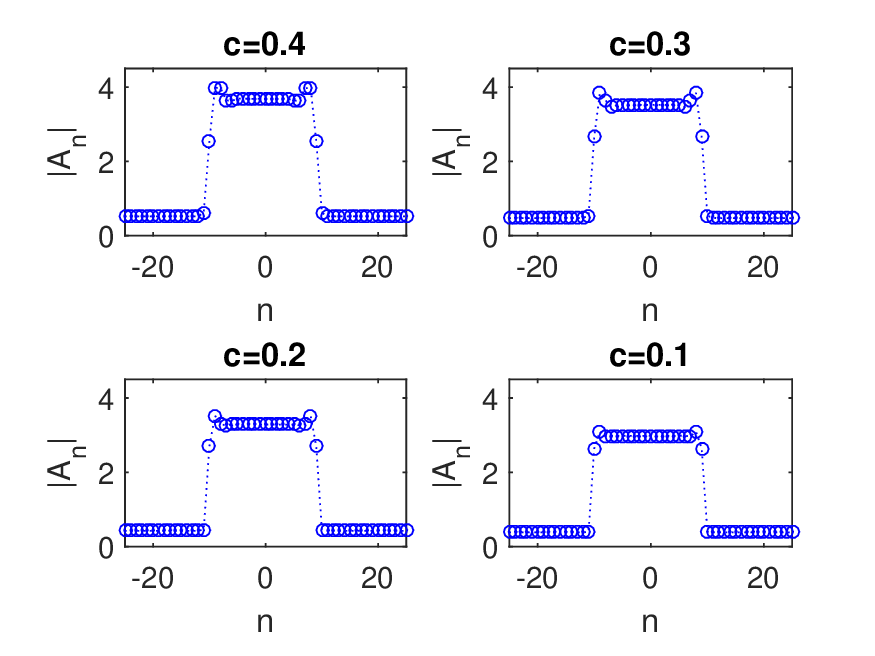}\label{subfig:prof_case_1_vary_c}}
		\caption{
			(a) The pinning region for case 1 and (b) {their} solution profiles at $P=4$ for different values of $c$.
			The blue dashed {lines indicate boundaries of} the pinning region.
			The red dotted-dashed line indicates the turning point of the uniform solution.
			The black dotted line indicates the one-active-site approximation of the pinning regions.
			{$\subset$-isolas formed at the right pinning region for $c\approx0.39$.}
		}
		\label{fig:all_pinning_cav1}
	\end{figure*} 
	\subsection{{$\subset$-shaped} isolas}\label{subsec:isola}
	%
	
	One interesting phenomenon observed in the snaking behavior of the optical cavities equation is that as the pinning region expands with decreasing $c$, the system eventually reaches a regime where the background (uniform state) of the localized solution ceases to exist. 
	
	This leads to an attachment or detachment process near the saddle-node bifurcations within the snaking branches. As a result, $\subset$-shaped isolas are formed when the control parameter $P$ is varied.
	
	Figure~\ref{fig:C_isolas_appear} illustrates the formation mechanism of these $\subset$-shaped isolas in case 1, as the coupling strength $c$ is varied. The isolas emerge around $c \approx 0.39$ for case 1, indicating a critical value of $c$ where the localized states detach from the main snaking structure and form isolated solution branches.

	\section{Pinning regions analysis}\label{sec:pinning_anal}

	%
	%
	%
	In this section, we analyze and discuss the pinning region and its approximation for both case 1 and case 2, focusing on the variation of the coupling strength $c$. This analysis provides insight into how the system's behavior changes as the coupling strength is adjusted and helps characterize the regions where localized solutions are pinned.
	
	\subsection{Pinning Region}\label{subsec:pinning}
	Pinning regions are bounded by turning points, which define the parameter values at which the stability of the localized solutions changes. These turning points mark the boundaries within which homoclinic snaking and localized structures can occur. To accurately compute the boundaries of the pinning regions, we solve the following extended system (see, e.g., \cite{govaerts2000numerical} for the details)
	\begin{equation}
		\left(
		\begin{array}{c}
			\partial_t A_n\\
			\mathcal{L}\varphi\\
			||\varphi||
		\end{array}
		\right)
		=\left(
		\begin{array}{c}
			0\\
			0\\
			1
		\end{array}
		\right),
		\label{eq:augemented_cavity}
	\end{equation}
	where $\varphi$ is an eigenvector of the Jacobian $\mathcal{L}$, corresponding to the largest eigenvalue, which becomes zero at the saddle-node bifurcation. 
	By utilizing Eq.~\eqref{eq:augemented_cavity}, one can compute the left and right boundaries of the pinning region without the need to calculate the entire snaking structure. However, obtaining accurate turning points, which define the boundaries of the pinning region, requires a well-chosen initial condition for the numerical continuation.
	It is also important to note that in the anti-continuum limit, the turning points of the uniform solutions continue to satisfy the extended system described by Eq.~\eqref{eq:augemented_cavity}. This ensures that the behavior near these points remains consistent, even as the coupling between sites weakens. This method enables us to locate the saddle-node bifurcations and determine the precise range of $P$ where localized solutions exist.

	
	Figures~\ref{fig:all_pinning_cav1} and \ref{fig:all_pinning_cav2} display the pinning regions for case 1 and case 2, respectively, as the coupling strength $c$ varies. In this analysis, we focus solely on the variation of $c$ to examine its effect on the pinning regions.	
	Figure~\ref{subfig:1_P_vs_c} shows the pinning region for case 1 as a function of $c$. As $c$ decreases, the pinning region expands, indicating that localized solutions are stable over a broader range of parameter values. Specifically, at $c \approx 0.39$, the right boundary of the pinning region reaches the right saddle point of the uniform solutions, leading to the formation of $\subset$-shaped isolas. 	
	This occurs because, at this critical value of $c$, the localized solution's background (uniform) state ceases to exist, causing the localized solutions around the right saddle point to detach from the snaking structure and form isolated solution branches (i.e., isolas). 	
	In contrast, as $c$ increases, the pinning region contracts. When $c$ becomes large, approaching the continuum limit, the on-site and off-site solutions converge and eventually merge, erasing the distinction between them.

Figure~\ref{fig:all_pinning_cav2} shows the pinning region for case 2 as the coupling strength $c$ is varied. The main distinction between case 2 and case 1 is that no $\subset$-shaped isolas are formed in case 2. 
This difference arises because, in case 2, the boundaries of the left and right pinning regions coincide with the saddle-node bifurcations of the uniform solution at $c = 0$ (decoupled limit). In contrast, for case 1, the right boundary of the pinning region reaches the saddle-node bifurcation around $c \approx 0.39$, leading to the formation of $\subset$-shaped isolas, as seen in Figs.~\ref{fig:all_pinning_cav1} and \ref{fig:all_pinning_cav2}. Therefore, the absence of isolas in case 2 is because the uniform background state persists over the entire range of $c$, preventing the detachment of localized solutions from the snaking branches.

\subsection{One-active-site approximation}\label{subsec:one_active}
As we vary the control parameter $P$, it becomes evident that there is effectively only one "active" node at the front connecting the two states of the uniform solutions, as depicted in Fig.~\ref{fig:onsite_offsite_several_P}. 
To simplify the analysis, we assume that only three nodes are involved in the dynamics as $P$ changes~\cite{kusdiantara2017homoclinic,susanto2018snakes}. These nodes are represented as:
\begin{equation}
	A_{n-1} = u_1, \quad A_n = \upsilon, \quad A_{n+1} = u_2,
	\label{eq:one_active_site_cav}
\end{equation}
where $u_1$ and $u_2$ correspond to the uniform background states on the upper and lower branches of the bifurcation diagram in Fig.~\ref{fig:case_unisol}, and $\upsilon$ is the active node connecting the two states.
By substituting Eq.~\eqref{eq:one_active_site_cav} into the time-independent equation \eqref{eq:cavity_ti}, we derive the following expression for the active node:
\begin{equation}
	F_a(\upsilon) := \delta \upsilon + \frac{\alpha|\upsilon|^2}{1+|\upsilon|^2} \upsilon + c(u_1 + u_2 - 2\upsilon) - P = 0.
	\label{eq:one_active_fun_cav}
\end{equation}

%
%

\begin{figure*}[t!]
	\centering
	\subfloat[\,Case 1]{\includegraphics[scale=0.45]{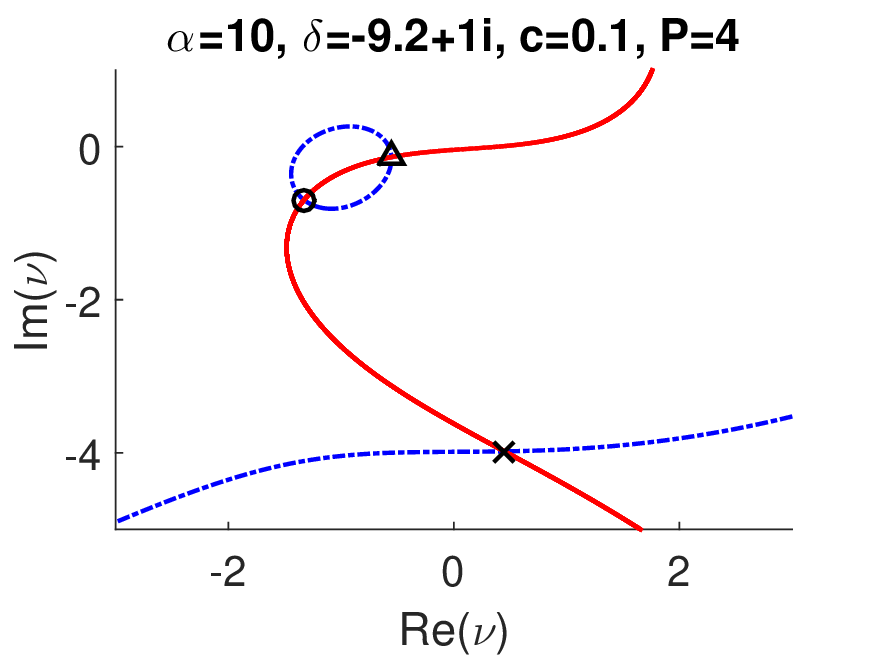}\label{subfig:cav_one_active1}}
	\subfloat[\,Case 2]{\includegraphics[scale=0.45]{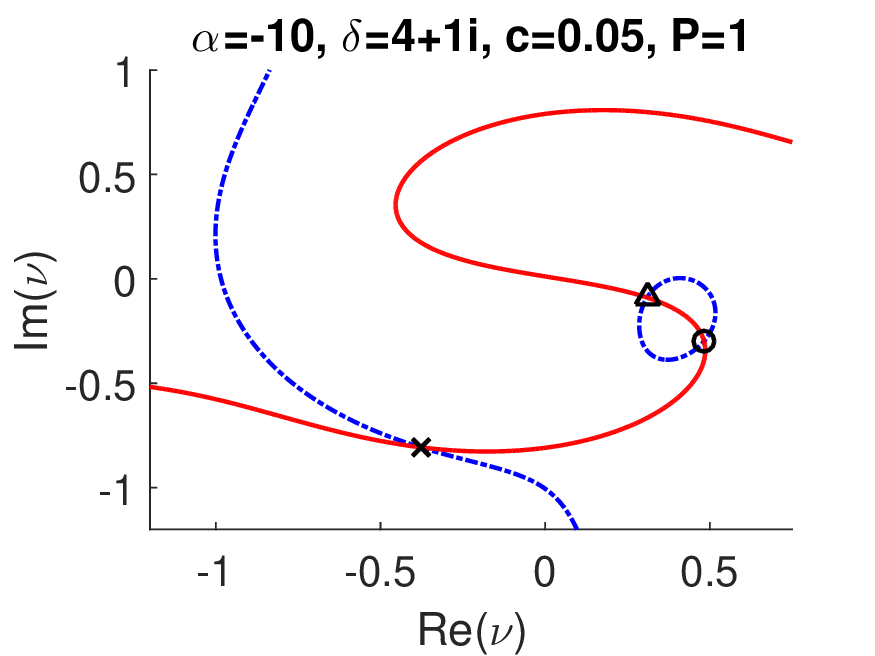}\label{subfig:cav_one_active2}}
	\caption{One-active-site function for cases 1 and 2. 
		The blue dotted-dashed line represents the real part of the one-active-site function $Re(F_a)$.
		The red solid line represents the imaginary part of the one-active-site function $Im(F_a)$. 
		Points $\textbigcircle$, $X$, and $\triangle$ represent the solutions of the one-active-site function \eqref{eq:one_active_fun_cav}.
	}
	\label{fig:cav_one_active}
\end{figure*}

For the one-active-site assumption to remain valid, the system's solutions must be sufficiently "discrete," meaning that the coupling is weak. In this regime, Eq.~\eqref{eq:one_active_fun_cav} can yield either one or three real solutions, which are associated with the snaking behavior. Two of these roots will vanish at a saddle-node bifurcation, which defines the boundary of the pinning region.
In Fig.~\ref{fig:cav_one_active}, the one-active-site function is shown for both case 1 and case 2. The three real roots are indicated in the figure. The boundaries of the pinning region are determined by the coalescence of these roots: specifically, when the points $\textbigcircle$ and $X$, or $\textbigcircle$ and $\triangle$, merge. The merging of these roots corresponds to the left and right boundaries of the pinning region, respectively. This collision mechanism is analogous to what is observed for the uniform solutions in Figs.~\ref{subfig:alpha_10_delta_-9_2_1i_contour} and \ref{subfig:alpha_-10_delta_4_1i_contour}.

\begin{figure}[t!]
	\centering
	\includegraphics[scale=0.6]{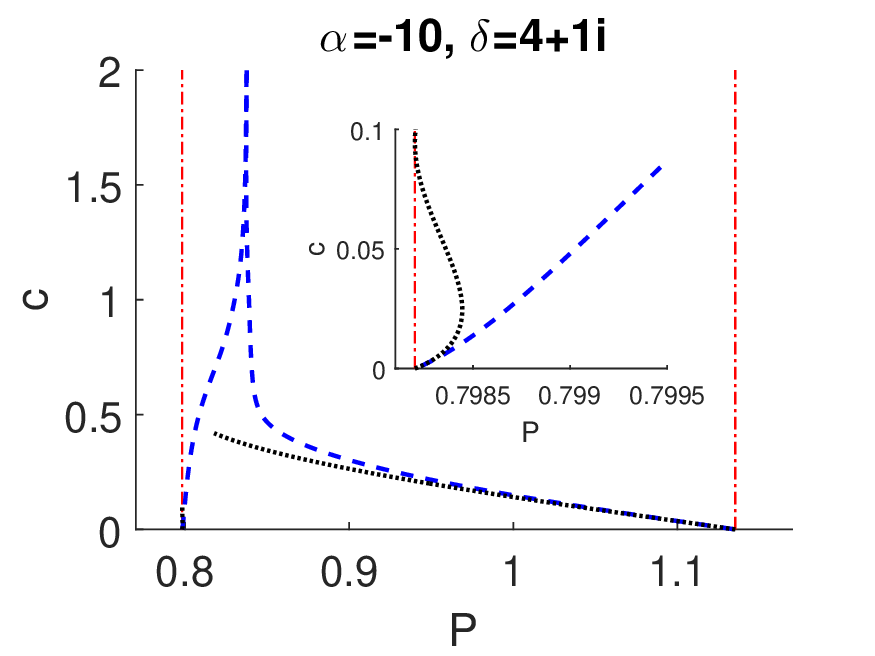}
	\caption{The same as {Fig}.\ \ref{subfig:1_P_vs_c}, but for case 2.
		{The inset compares the numerics and the one-active-site approximation of the left pinning region boundary.}
		{There are no $\subset$-shaped isolas for this case because the background state of the localized solution always exists for any value of $c$.}}
	\label{fig:all_pinning_cav2}
\end{figure}

Figures~\ref{fig:all_pinning_cav1} and \ref{fig:all_pinning_cav2} compare the numerical results from Eq.~\eqref{eq:cavity_ti} with the predictions made by the one-active-site approximation, Eq.~\eqref{eq:one_active_fun_cav}. Figure~\ref{subfig:1_P_vs_c} shows that the one-active-site approximation works well for small values of $c$. However, for larger $c$, the solution profile develops more than a single active node in the "upper" state, as seen in Fig.~\ref{subfig:prof_case_1_vary_c}. In this case, the one-active-site assumption breaks down and is no longer valid.

\begin{figure*}[t!]
	\centering
	\includegraphics[scale=0.5]{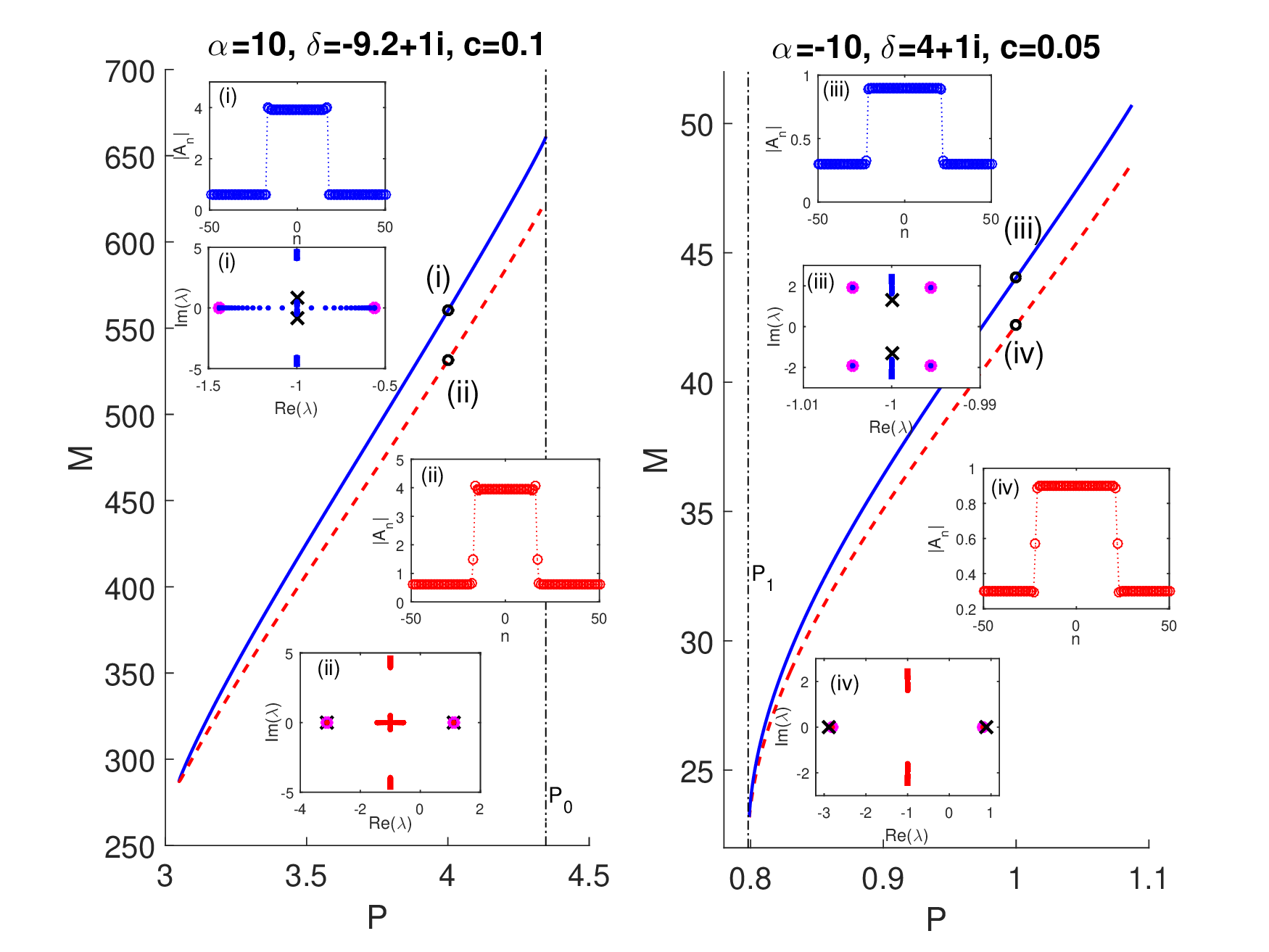}
	\caption{
		{Left and right panels are bifurcation diagrams for cases 1 and 2, respectively. The insets show (in circles) localized solutions and their spectrum in the complex plane, obtained from solving Eqs.\ \eqref{eq:cavity_ti} and \eqref{eq:evp_cav}, respectively. Magenta circles show the maximum and minimum real part of the spectra. Crosses show our approximation from Eq.\ \eqref{eq:one-evp_cav} with \eqref{eq:one_active_fun_cav}.}
	}
	\label{fig:spectra_cav}
\end{figure*}
The comparison between the numerical results and the one-active-site approximation for case 2 is presented in Fig.~\ref{fig:all_pinning_cav2}. The inset in Fig.~\ref{fig:all_pinning_cav2} highlights a key aspect of this comparison, particularly focusing on the boundary of the left pinning region.
The simulation results demonstrate that the one-active-site approximation performs well when the coupling strength $c$ is small. Moreover, the approximation provides better accuracy when the pinning region boundary is located relatively far from the turning point of the uniform solution. 
In general, the one-active-site approximation yields reliable results when the ``upper" and ``lower" states of the localized solutions are relatively flat (uniform) or weakly coupled, and when only a single active node mediates the connection between them. Under these conditions, the approximation effectively captures the system's behavior.

Next, the critical eigenvalue of the localized solutions in the pinning region can be approximated using the one-active-site model. This is done by considering the dynamics of Eq.~\eqref{eq:one_active_fun_cav}, i.e.,
\begin{equation}
	-i \, \upsilon_t = F_a(\upsilon),
	\label{eq:one-cav_evp}
\end{equation}
where $\tilde{\upsilon} = \tilde{\upsilon}_R + i \tilde{\upsilon}_I$ is a solution to the one-active-site function in Eq.~\eqref{eq:one_active_fun_cav}.
Linearizing Eq.~\eqref{eq:one-cav_evp} around $\tilde{\upsilon}$ by writing $\upsilon = \tilde{\upsilon} + \epsilon (\upsilon_r + i \upsilon_i) e^{\lambda t}$ for small $|\epsilon|$ yields the following eigenvalue problem:
\begin{equation}
	\lambda
	\begin{pmatrix}
		\upsilon_r \\
		\upsilon_i
	\end{pmatrix}
	=
	\begin{pmatrix}
		s_{11} & s_{12} \\
		s_{21} & s_{22}
	\end{pmatrix}
	\begin{pmatrix}
		\upsilon_r \\
		\upsilon_i
	\end{pmatrix},
	\label{eq:one-evp_cav}
\end{equation}
where
\begin{equation*}
	\begin{array}{ccl}
		s_{11} &=& -\delta_i - m_{11}(\tilde{\upsilon}_R, \tilde{\upsilon}_I), \\
		s_{12} &=& -\delta_r - 2c - m_{12}(\tilde{\upsilon}_R, \tilde{\upsilon}_I), \\
		s_{21} &=& \delta_r + 2c - m_{21}(\tilde{\upsilon}_R, \tilde{\upsilon}_I), \\
		s_{22} &=& -\delta_i - m_{22}(\tilde{\upsilon}_R, \tilde{\upsilon}_I).
	\end{array}
\end{equation*}
Here, $m_{11}$, $m_{12}$, $m_{21}$, and $m_{22}$ are defined in Eq.~\eqref{eq:m_eq_cav}. From this, we can derive the eigenvalues as:
\begin{equation}
	\lambda(P) = \frac{1}{2} \left( s_{11} + s_{22} \pm \sqrt{(s_{11} - s_{22})^2 + 4s_{12}s_{21}} \right).
	\label{eq:dis_rel_oneactive_cav}
\end{equation}

The inset in Fig.~\ref{fig:spectra_cav} presents the numerically computed spectrum of the localized solutions alongside the results from our one-active-site approximations. A good comparison between the two shows that the approximation performs well in most cases.
However, the critical eigenvalues predicted by our approximation do not match the numerical results when the eigenvalues are associated with the background state rather than the front of the localized solution. This discrepancy arises because the one-active-site approximation is specifically designed to capture the dynamics of the front, or active site, of the localized solution, as highlighted in the insets (i) and (iii) of Fig.~\ref{fig:spectra_cav}. 
In these cases, the background state plays a more significant role in determining the stability of the localized solution than the front itself. Therefore, while the one-active-site approximation is a reliable tool for assessing the stability of localized solutions in weakly coupled systems, it is less accurate when the stability is governed by the background states.

\section{Conclusion}\label{sec:conclusion}

In this paper, we have demonstrated for the first time how the one-active-site approximation can be used to qualitatively and quantitatively analyze the shape of the pinning region in spatially discrete pattern-forming systems in the weak coupling limit. Furthermore, we have shown that this method approximates localized solutions and their critical eigenvalues in this regime.

We applied this method to a previously studied model of discrete optical cavities with saturable nonlinearity and introduced several novel results for this system. In particular, we demonstrated how the bifurcations of localized states from the uniform state depend on the coupling strength. Through numerical continuation, we computed the pinning regions within which homoclinic snaking can occur. Additionally, we numerically identified parameter regions where the snaking structure breaks up into a series of $\subset$-shaped isolas in the bifurcation diagrams.

The one-active-site approximation has potential applications beyond the specific system studied in this paper. For example, applying this method to similar equations with cubic nonlinearity, such as the discrete Lugiato-Lefever equation~\cite{peschel2004discrete,egorov2007does,egorov2013spontaneously,averlant2017coexistence}, could yield further insights. Moreover, the recent rigorous work by Bramburger and Sandstede~\cite{bramburger2020localized,bramburger2021isolas} on the existence and stability of localized patterns in spatially discrete versions of the bistable Ginzburg-Landau equation demonstrates the usefulness of such approximations in the anti-continuum limit.

We also note that the one-active-site approximation can be extended to higher spatial dimensions, unlike traditional spatial dynamics methods often used to describe homoclinic snaking~\cite{kusdiantara2019snakes,kusdiantara2022snakes}. Furthermore, it would be interesting to explore the existence of multipulse solutions on isolas, as discussed in~\cite{bramburger2021isolas}. We are also interested in extending the approach of one-active-site approximation to multi-site approximation. While the number of equations to be considered will increase, such an extension will provide a larger validity interval of the coupling constant. 
{The existence of localized solutions to the model considered in this report, without applied optical pump and loss terms (i.e., $P=\text{Im}(\delta)=\text{Im}(\alpha)=0$), was proven rigorously in \cite{pankov2007gap,pankov2008periodic}. The extension to the general case is still open.}

\section*{Data availability}
No data was used for the research described in the article.

\section*{Declaration of competing interest} 
The authors declare that they have no known competing financial interests or personal relationships that could have appeared to influence the work reported in this paper.

\section*{CRediT authorship contribution statement}
The manuscript was written with contributions from all authors. All authors have given their approval to the final version of the manuscript.\\

\textbf{RK}: Software, Validation, Formal Analysis, Investigation, Writing - Original Draft; \textbf{HS}: Conceptualization, Methodology, Validation, Writing - Original Draft, Writing - Review \& Editing; \textbf{ARC}: Conceptualization, Methodology, Validation, Writing - Review \& Editing.

\section*{Acknowledgements}
RK acknowledged support through Hibah PPMI FMIPA ITB 2024 (617I/IT1.C02/KU/2024) by Institut Teknologi Bandung. 
HS acknowledged support by Khalifa University through a Competitive Internal Research Awards Grant (No.\ 8474000413/CIRA-2021-065) and Research \& Innovation Grants (No.\ 8474000617/RIG-S-2023-031 and No.\ 8474000789/RIG-S-2024-070).

\bibliographystyle{elsarticle-num}

\end{document}